\documentclass[twocolumn,showpacs,preprintnumbers,aps,prd,a4paper,superscriptaddress,nofootinbib,amsmath,amssymb,floats,floatfix]{revtex4-1}

\usepackage[usenames,dvipsnames,svgnames,table]{xcolor}
\usepackage[colorlinks=true,linkcolor=MidnightBlue,citecolor=RawSienna,urlcolor=MidnightBlue]{hyperref}
\usepackage{graphicx}
\usepackage{natbib}
\usepackage{lmodern}

\usepackage{amsfonts}
\usepackage{amssymb}
\usepackage{amsmath}

\usepackage{bm}

\usepackage{feynmp}

\RequirePackage{ifpdf}
\ifpdf
\else 
\fi

\renewcommand{\d}{\mathrm{d}}

\newcommand{\LambdaUV}{\Lambda_{\mathrm{UV}}}
\newcommand{\LambdaSM}{\Lambda_{\mathrm{IR}}}
\newcommand{\tildeLambdaSM}{\tilde{\Lambda}_{\mathrm{IR}}}

\newcommand{\Damp}{D_{\text{amp}}}
\newcommand{\Dsub}{D_{\text{sub}}}
\newcommand{\MSM}{M_{\mathrm{SM}}}
\newcommand{\tildeMSM}{\tilde{M}_{\mathrm{SM}}}
\newcommand{\Hage}{H_{\text{age}}}

\newcommand{\mtop}{m_{\mathrm{t}}}
\newcommand{\tildemtop}{\tilde{m}_{\mathrm{t}}}

\newcommand{\eV}{\mathrm{eV}}

\newcommand{\GeV}{\mathrm{GeV}}
\newcommand{\TeV}{\mathrm{TeV}}

\newcommand{\Mp}{M_{\mathrm{P}}}
\newcommand{\Lmatter}{\mathcal{L}_{\mathrm{m}}}
\newcommand{\Smatter}{S_{\mathrm{m}}}

\newcommand{\GaussBonnet}{\mathrm{GB}}
\newcommand{\RGaussBonnet}{R_{\GaussBonnet}}

\newcommand{\leftaverage}{\langle\hspace{-0.25em}\langle}
\newcommand{\rightaverage}{\rangle\hspace{-0.25em}\rangle}
\newcommand{\STaverage}[1]{\leftaverage{{#1}}\rightaverage}

\newcommand{\semibold}[1]{{\fontseries{b}\selectfont{#1}}}

\newcommand{\para}[1]{\par\vspace{2mm}\noindent\semibold{{#1.}---}\ignorespaces}
\newcommand{\varpara}[1]{\par\vspace{2mm}\noindent\semibold{{#1}---}\ignorespaces}

\DeclareMathOperator{\Or}{O}

\begin{document}
\begin{fmffile}{diags}
\title{Ultraviolet sensitivity of the cosmological sequester}
\author{Michaela~G.~Lawrence}
\email{M.G.Lawrence@sussex.ac.uk}
\affiliation{Astronomy Centre, Department of Physics and Astronomy, University of Sussex, Brighton, BN1 9QH, United Kingdom}
\author{David~Seery}
\email{D.Seery@sussex.ac.uk}
\affiliation{Astronomy Centre, Department of Physics and Astronomy, University of Sussex, Brighton, BN1 9QH, United Kingdom}

\begin{abstract}
We revisit the ``sequester'' proposal of Kaloper, Padilla and collaborators,
in which the amplitude of the cosmological
constant is decoupled
from large contributions due to loops containing
Standard Model particles.
We review the different formulations of the model that have
appeared in the literature, and
estimate the importance of a particular class of
quantum corrections---those that dress the interaction between
the ``rigid'' scalars
and infrared properties of the spacetime such as its 4-volume
and integrated curvature.
In formulations that do not adequately sequester graviton loops
we argue that dressing of these interactions
causes further failures of complete sequestration.
We estimate the size of the effect and find that it is typically
smaller than the cosmological term directly induced by
loops containing a single virtual graviton.
Meanwhile, in the most developed formulation of the scenario
(where a rigid scalar couples to the Gauss--Bonnet density),
this dressing can be absorbed into a rescaling
of the rigid fields
and is therefore harmless.
\end{abstract}

\maketitle

\section{Introduction}
It is nearly 40 years since the cosmological
constant problem
was first stated 
clearly~\cite{Wilczek:1983as,Weinberg:1988cp}.
(For the earlier history, see Ref.~\cite{Rugh:2000ji,Straumann:2002tv}.)
Despite immense efforts over the intervening decades,
it 
remains the most enigmatic
component of the concordance
cosmological model.
The problem is simple to state.
Observation requires
the cosmological constant $\Lambda$ to dominate the present
Hubble rate,
and
therefore $3H_0^2 \Mp^2 \approx \Lambda$. 
The measured value of $H_0$ gives
an estimate
$\Lambda \approx 10^{-12} \, \eV^4$. 
Meanwhile quantum contributions to
$\Lambda$ from Standard Model particles are
much larger.
Why, then, is the measured value so small?

\para{The case for `fine tuning'}
The operational meaning of
$\Lambda$ is less clear than other quantities
that are known to receive large quantum corrections,
such as the running couplings
that appear in scattering amplitudes,
because it couples only at wavenumber zero
where scattering does not occur.
Nevertheless,
like any low-energy constant,
$\Lambda$ presumably can be divided into an incalculable
ultraviolet contribution
$\LambdaUV$
from unknown physics lying above the Standard Model,
and an infrared contribution
$\LambdaSM$ generated by quantum corrections with
Standard Model particles running in the loops.
We expect $\LambdaSM \sim \mtop^4$ 
from loop diagrams containing the top quark, which is the
heaviest Standard Model particle.%
    \footnote{It is often said that the low-energy calculation
    yields $\LambdaSM \sim \Mp^4$, but this is not justified. 
    Although the vacuum energy computed with a hard momentum
    cutoff is quartically divergent, it must be remembered that
    \emph{cutoffs do not track dependence on
    heavy masses}~\cite{Burgess:1992gx}.
    The low-energy theory cannot yield a
    trustworthy dependence on any mass
    scale heavier than it contains itself, and the heaviest
    mass described by the effective Lagrangian for the Standard Model
    is the top mass $\mtop$.
    See also {\S}4.2 of Ref.~\cite{Lyth:2010ch},
    and Refs.~\cite{Sloth:2010ti,Maggiore:2010wr,Mangano:2010hw},
    which show explicitly that the quartically divergent
    terms cannot be interpreted as a dark energy component.
    
    If there \emph{is} a contribution to the cosmological constant
    of order $\sim \Mp^4$, it comes from $\LambdaUV$ and not $\LambdaSM$. 
    For example, this might happen if local field theory remains valid all
    the way up to the Planck scale, and the low-energy gravitational force
    is generated by integrating out one or more particles of mass $\sim \Mp$.
    But the outcome could be different
    if local field theory fails as a good approximation to Nature
    at some much lower scale.}

It follows that
$\Lambda = \LambdaUV + \LambdaSM$ should 
be of order $\mtop^4 \sim (175\, \GeV)^4$ or larger 
unless $\LambdaUV$ is accurately balanced to cancel large contributions 
from $\LambdaSM$. 
The measurement $\Lambda \sim 10^{-12} \, \eV^4$ apparently
implies that
$\LambdaUV$ is balanced so that cancellation occurs to
roughly 56 decimal places.
    \footnote{The large number of decimal places required is because 
    cancellation has to occur in $\LambdaUV + \LambdaSM$.} 
The scales that contribute to $\LambdaUV$ and $\LambdaSM$
are very different, so there is no reason
why $\LambdaUV$ should be related to Standard Model energies.
This makes it unlikely that
cancellation happens by accident.

Unless new physics changes the relationship
between $\Lambda$, $\LambdaSM$ and $\LambdaUV$,
the most plausible conclusion is that
whatever
determines $\LambdaUV$
must be
constrained by some principle
forcing $\Lambda$ to be nearly zero.
Such a principle would strongly violate decoupling,
because it would make the Wilson
coefficients of the low-energy action
into highly sensitive functions of the ultraviolet
boundary condition.
The apparent tuning we observe would be a consequence
of this exquisite sensitivity.

It is certainly possible
that the correct resolution of the cosmological
constant problem
involves a failure of decoupling along these lines.
Unfortunately,
modern ideas in particle physics
have not yielded any candidate principle that could be
responsible for the smallness of $\Lambda$.
Moreover, the failure of decoupling makes
robust low-energy model building difficult.
For these reasons it is now more common to look
for an alternative resolution.

\para{Overview of this paper}
In this paper we revisit the ``sequester'' proposal of Kaloper
\& Padilla~\cite{Kaloper:2013zca,Kaloper:2014dqa,Kaloper:2015jra,Kaloper:2016jsd}.
This is a concrete scenario for new physics that
changes the argument given above
by removing
(``sequestering'')
the low-energy contribution from all Standard Model
particles.
The outcome is that
the observed cosmological constant $\Lambda$ would be
set by $\LambdaUV$, unless there are further contributions from
new unsequestered sectors.

By itself the sequester
(or at least its simplest versions)
would not explain the observed magnitude
of $\Lambda$.%
    \footnote{In {\S}\ref{sec:omnia-sequestra} we will see
    that the most developed version of the sequester would
    absorb $\LambdaUV$ in addition to $\LambdaSM$,
    at the cost of introducing a new cosmological-like term
    associated with an unknown scale $\mu$.
    See Eqs.~\eqref{eq:omnia-sequestra-Einstein-eqn}
    and~\eqref{eq:omnia-sequestra-residual}.
    Therefore, no matter what strategy we choose, it seems that
    one can not arrive at an unambiguous prediction
    for the observed
    value of $\Lambda$.}
Even if all matter species participate in sequestration,
there would still be a puzzle
if we expect $\LambdaUV$ to receive contributions
larger than $10^{-3} \, \eV$.
This might be the case, for example, if
low-energy Einstein gravity is an effective description
generated by integrating out one or more
Planck-mass particles.
The advantage of the sequester
is that
the small observed value
no longer requires cancellations
between $\LambdaUV$ and $\LambdaSM$.
We express this by saying that its value is
\emph{technically natural}
within the Standard Model.
Whether or not it is technically natural
with respect to the ultraviolet model is a question
that can be resolved only when that theory is
specified.

The status of arguments based on technical naturalness
has been called into question following
the discovery of a Higgs particle at $M \sim 125 \, \text{GeV}$
without new accompanying particles~\cite{Richter:2006gsw,Hossenfelder:2018jew}.
In the formulation we are using, ``naturalness''
has a clear meaning in terms of sensitivity---or lack of it---%
to large corrections between widely separated
scales~\cite{Giudice:2013yca,Williams:2015gxa,Giudice:2017pzm}.%
	\footnote{This is a broader definition than the original
	concept of technical naturalness due to t'Hooft~\cite{tHooft:1979rat}.
	t'Hooft's criterion that a small parameter $y$ is natural
	if the symmetry of the theory is enlarged in the limit $y \rightarrow 0$
	is a sufficient but not necessary condition for widely separated
	scales to decouple in this sense.}
This is not merely an aesthetic choice,
and
accordingly Nature may or may not be ``natural''
in our sense.
Nevertheless,
it is reasonable to expect this concept
of naturalness
to be a useful guide because experience has shown
that the vast majority of physical phenomena \emph{do}
decouple in this way.

Clearly
we should not be satisfied with a theory in which $\Lambda$
is made technically natural
at the expense of other low-energy
constants that receive large ultraviolet corrections.
If this occurs we have not removed the problem,
but merely translated it
from one low-energy sector to another.
In this paper we aim to apply this test to the
sequester model.

\para{Synopsis}
Two principal variants of the sequester
have been discussed in the literature.
In the first version, one works in the Einstein frame
and couples the sequestered sectors to a conformally
rescaled metric.
This version
was introduced in Refs.~\cite{Kaloper:2013zca,Kaloper:2014dqa};
see Ref.~\cite{Padilla:2015aaa} for a pedagogical
description.
We describe it as the `Einstein frame model'.
The conformal rescaling dynamically
adjusts mass scales in the sequestered sector
relative to the fixed Planck scale.
A global constraint couples the cosmological term
to this conformal factor, allowing it to absorb
contributions from pure matter loops.
In this version,
loops involving virtual gravitons are known to reintroduce
unsequestered corrections to the observed $\Lambda$.
We discuss this model and the degree to which it ameliorates
ultraviolet sensitivity of the cosmological constant
in {\S}\ref{sec:sequester-model1}.

A second variant was introduced in
Refs.~\cite{Kaloper:2015jra,Kaloper:2016jsd}.
In this version one works in the Jordan frame
and there is no auxiliary rescaled metric.
There are two global constraints,
the first of which couples the gravitational
scale to the mean Ricci curvature of spacetime.
The second couples the cosmological constant to the
total spacetime volume
and a physical mass scale $\mu$,
which is \emph{a priori} unknown.
The is the `Jordan frame model.'
In this version one can adjust the way in
which the global constraints
couple to
spacetime curvature so that loops of virtual gravitons
are also absorbed.
This version of the sequester
and its ultraviolet properties
are discussed in {\S}\ref{sec:sequester-model2}.
We conclude in {\S}\ref{sec:conclusions}.

\para{Notation}
We work in units where $c = \hbar = 1$.
The (reduced) Planck mass is
$\Mp \equiv (8\pi G)^{-1/2} = 2.435 \times 10^{18} \, \GeV$.
We express the cosmological constant in terms of an energy
scale $\Lambda$ with engineering dimension $[\mathrm{M}]^4$.
The corresponding ``cosmological term'' in the Einstein equations
is $\Lambda$.
We generally frame our calculations in Minkowski space to avoid
unneeded complexities associated with curved spacetime;
because ultraviolet properties do not depend on these curvature
scales, this procedure does not forfeit any
essential generality.

\section{Einstein frame model}
\label{sec:sequester-model1}

\subsection{The sequester action}
In this section we
briefly review the sequester mechanism in
Einstein frame~\cite{Kaloper:2013zca,Kaloper:2014dqa},
and discuss its
ultraviolet sensitivity.
The gravitational action is written in
terms of the Einstein-frame metric $g_{\mu\nu}$.
Sequestration of one or more
matter sectors is achieved
by coupling them to a conformally rescaled
(`Jordan-frame')
metric $\tilde{g}_{\mu \nu} = \lambda^2 g_{\mu \nu}$, viz.
\begin{equation}
\begin{split}
    S = & \int \d^4 x \; \sqrt{-g} \Bigg( \frac{\Mp^2}{2} R(g)
    - \Lambda
    + \LambdaUV
    - \lambda^4 \Lmatter(\tilde{g}^{\mu\nu}, \Psi) \Bigg)
    \\ &
    \mbox{} + \sigma\left( \frac{\Lambda}{\lambda^4 \mu^4} \right) .
\end{split}
\label{eq:model-one}
\end{equation}
If multiple sectors are to be sequestered their actions
should appear additively.
In Eq.~\eqref{eq:model-one},
$R(g) = g^{\mu\nu} R_{\mu\nu}(g)$ is the Ricci scalar
constructed using the Einstein-frame metric $g_{\mu\nu}$,
$\Lmatter$ is a matter Lagrangian density,
and $\Psi$ schematically stands for the different species of
sequestered matter fields.
We assume these to be the Standard Model fields.
The low energy contribution to the cosmological
constant, $\LambdaSM$, does not appear in
Eq.~\eqref{eq:model-one}
explicitly.
It is generated by
the infrared part of loop corrections to
$\Lmatter$.
The bare cosmological constant (if there is one),
plus any contributions from unsequestered sectors that
have been integrated out to produce~\eqref{eq:model-one},
are included in $\LambdaUV$.

The quantity $\Lambda$ is no longer the
combination $\LambdaUV + \LambdaSM$,
but is rather a new field
that can loosely be regarded as a counterterm
for $\LambdaSM$.
In particular, although it participates in the
path integral,
we take $\Lambda$ to have no local
degrees of freedom.
It is determined classically by
extremization of the action.
The dimensionless conformal rescaling $\lambda$ is taken to be a
field
of the same kind.

The global term $\sigma$ is a function of $\Lambda$ and $\lambda$
in the specific dimensionless combination $\Lambda / (\lambda \mu)^4$.
Critically, it does \emph{not} couple to either
the Einstein- or Jordan-fame metric,
and therefore does not source the global gravitational field.
The
scale $\mu$ has dimension $[\mathrm{M}]$,
but its precise meaning
depends on the definition of $\sigma$.
We will discuss its significance in more detail
in~{\S}\ref{sec:model-two} below.
Finally, for reasons to be explained below, we should take $\sigma$
to be an odd function of its argument.
It is otherwise assumed to be an arbitrary smooth function.

The rigidity of $\Lambda$ and $\lambda$ is unusual, but can be given
a local, microscopic basis
in terms of integrals of a four-form flux $F_4$ over spacetime~\cite{Kaloper:2015jra}.
Such a flux is a top-order form in $d=4$ dimensions and therefore
acts as a volume form in the integral $\int F_4$.
In particular, this integral can be written
without requiring a metric.
Borrowing terminology from thermodynamics, we describe terms such
as $\int F_4$ that do not scale with $g_{\mu\nu}$ as
\emph{intensive}.
Ordinary contributions to the action such as
$\int (\star 1)$
are conversely \emph{extensive}.
Notice that if $\sigma$ does not scale at least
with the coordinate volume of spacetime,
this violates
Hawking's suggestion that the action
should be additive over cobordant regions
in order for quantum gravitational
amplitudes to superpose correctly~\cite{Hawking:1979ig}.

\subsection{Low-energy phenomenology}
\label{sec:low-energy-model-one}
We now consider low-energy solutions to~\eqref{eq:model-one}.
First, notice that
the matter contribution in~\eqref{eq:model-one}
can be written%
	\footnote{We take this as the definition of
	the matter action $\Smatter$.}\begin{equation}
    \Smatter \equiv -\int \d^4 x \, \sqrt{-\tilde{g}} \;
    \Lmatter(\tilde{g}^{\mu\nu}, \Psi) .
\end{equation}
Therefore it is clear that the matter fields
$\Psi$
are minimally coupled to
the Jordan-frame metric $\tilde{g}_{\mu\nu}$.
By taking $\tilde{g}_{\mu\nu}$ to be flat up to corrections from the
Newtonian potential,
it follows that
predictions for laboratory measurements
in a weak gravitational field will
match those of the unsequestered Standard Model.

We conclude that
the masses and other properties of the Standard Model
reported by the Particle Data Group~\cite{Tanabashi:2018oca}
are those measured in $\tilde{g}_{\mu\nu}$.
We denote these experimental scales with a tilde, viz.
$\tilde{M}_Z$, $\tildemtop$.
They are related
to scales measured in the metric $g_{\mu\nu}$
by a conformal rescaling
$\tilde{M}_Z \rightarrow M_Z = \lambda \tilde{M}_Z$.

\para{Sequestering low-energy loops}
Extremization of~\eqref{eq:model-one} with respect to $\Lambda$
and $\lambda$ yields,
\begin{subequations}
\begin{align}
    \label{eq:model1-Lambda-eqn}
    \frac{\sigma'}{(\lambda \mu)^4}
    &
    = \int \d^4 x \; \sqrt{-g}, \\
    \label{eq:model1-lambda-eqn}
    4 \frac{\Lambda}{(\lambda \mu)^4} \sigma'
    & 
    = \int \d^4 x \; \sqrt{-\tilde{g}} \, {{\tilde{T}^\mu}}_{\hspace{2mm}\mu}
    = \int \d^4 x \, \sqrt{-g} \, {T^\mu}_\mu
    ,
\end{align}
\end{subequations}
where a prime $'$ denotes differentiation of $\sigma$
with respect to its argument, and the
Jordan-frame
energy--momentum tensor $\tilde{T}_{\mu\nu}$
measured with respect to $\tilde{g}^{\mu\nu}$
is defined by
\begin{equation}
    \tilde{T}_{\mu\nu} \equiv -\frac{2}{\sqrt{-\tilde{g}}}
    \frac{\delta \Smatter}{\delta \tilde{g}^{\mu\nu}} .
    \label{eq:energy-momentum-tensor-def}
\end{equation}
A similar definition applies
for the Einstein-frame energy-momentum tensor $T_{\mu\nu}$,
which is measured with respect to $g^{\mu\nu}$.
The two definitions are related by
$\tilde{T}_{\mu\nu} = \lambda^{-2} T_{\mu\nu}$. 
We assume $\sigma' \neq 0$
at the extremum.
To allow consistent solutions with $\Lambda < 0$
but $\lambda > 0$ we require $\sigma'(x)$ to be even,
and hence $\sigma(x)$ must be odd,
as stated above.

Taking the ratio of Eqs.~\eqref{eq:model1-lambda-eqn}
and~\eqref{eq:model1-Lambda-eqn}
yields a constraint for $\Lambda$,
\begin{equation}
    \Lambda = \frac{1}{4}
    \STaverage{{T^\mu}_\mu} ,
    \label{eq:Lambda-soln}
\end{equation}
where $\STaverage{\cdots}$
denotes \emph{spacetime} averaging
in the metric $g_{\mu\nu}$,
i.e.
$\STaverage{Q} \equiv \int \d^4 x \, \sqrt{-g} \, Q / \int \d^4 x \, \sqrt{-g}$.
Since we assume $\sigma$ is differentiable,
Eq.~\eqref{eq:model1-Lambda-eqn}
requires the volume of spacetime to be finite if
we wish to avoid $\lambda = 0$.
(This would conformally rescale all masses in the sequestered
sector to zero.)
It follows that the spacetime average
$\STaverage{Q}$ can be defined, even
if it is difficult to evaluate in practice.

\begin{figure}
	\begin{center}
    		\includegraphics[width=0.85\columnwidth]{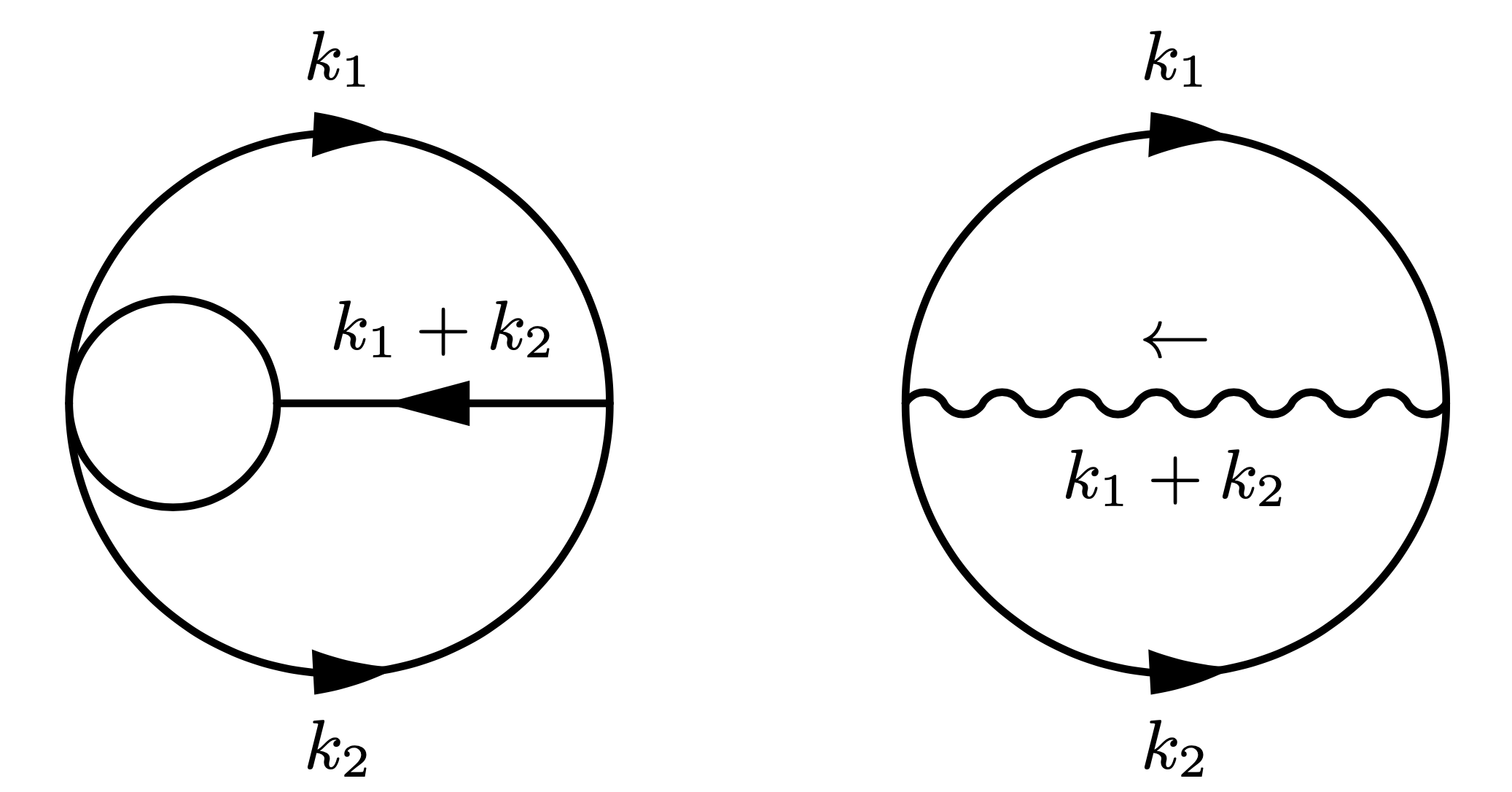}
	\end{center}
	\caption{\label{fig:cc-loops}Diagrams contributing to renormalization of the
	cosmological constant in the
	Einstein frame. Solid lines represent
	generic Standard Model particles,
	and wiggly lines represent
	spin-2 excitations of the metric $g_{\mu\nu}$.
	\semibold{Left}:
	pure Standard Model loop.
	This diagram scales like $\lambda^4$ when expressed in 
	terms of experimentally-measured mass scales,
	and is captured by the sequester.
	\semibold{Right}:
	mixed Standard Model and graviton loop.
	Because the Einstein-frame graviton propagator
	is proportional to the hard
	scale $\Mp^{-2}$,
	this diagram must scale like $\lambda^6$ rather
	than $\lambda^4$.
	As explained in the main text, this implies it
	is not captured by the sequester in Einstein frame.}
\end{figure}

The Einstein equation that follows from~\eqref{eq:model-one}
is
\begin{equation}
\begin{split}
	\Mp^2 G_{\mu\nu}
	& =
	T_{\mu\nu} - (\Lambda - \LambdaUV) g_{\mu\nu}
	\\
	& =
	T_{\mu\nu} - \frac{1}{4} \STaverage{{T^\mu}_\mu} g_{\mu\nu} + \LambdaUV g_{\mu\nu} ,
\end{split}
\label{eq:sequester-Einstein}
\end{equation}
where $G_{\mu\nu}(g) = R_{\mu\nu}(g) - R(g) g_{\mu\nu}/2$ is the usual Einstein tensor
constructed from $g_{\mu\nu}$.
As explained above, the $\sigma$ term in the action
does not couple to $g_{\mu\nu}$ and therefore does not source
a long-wavelength gravitational field.
Diffeomorphism invariance guarantees that any matter loops
renormalize the cosmological term in $\Lmatter$
measured using $\tilde{g}_{\mu\nu}$ (see Fig.~\ref{fig:cc-loops}), and therefore
\begin{equation}
    \tilde{T}_{\mu\nu} = \tildeLambdaSM \tilde{g}_{\mu\nu}
    + \tilde{\tau}_{\mu\nu}( \tilde{g}, \Psi, \cdots) ,   
\end{equation}
where the `subtracted' energy--momentum tensor
$\tilde{\tau}_{\mu\nu}(\tilde{g}, \Psi, \ldots)$
vanishes outside matter.
We have added a tilde to $\LambdaSM$ to indicate that it is built
from scales such as
$\tilde{m}_{\text{t}}$
measured in a homogeneous gravitational field.
When expressed in terms of $T_{\mu\nu}$ we obtain
\begin{equation}
	T_{\mu\nu} = \lambda^4 \tildeLambdaSM g_{\mu\nu}
	+ \lambda^2 \tilde{\tau}_{\mu\nu}( \tilde{g}, \Psi, \ldots) ,
\end{equation}
It follows that the effective Einstein equation can be written
\begin{equation}
	\Mp^2 G_{\mu\nu} =
	\tau_{\mu\nu}
	- \frac{1}{4} \STaverage{{\tau^\rho}_\rho} g_{\mu\nu}
	+ \LambdaUV g_{\mu\nu}
	.
	\label{eq:model1-seq-Einstein-eqn}
\end{equation}
The conclusion is that,
in the Einstein equation,
the low-energy contribution
 $\LambdaSM$
is removed to all orders in the loop expansion
of $\Lmatter$.

\varpara{What has been achieved?}
To reiterate, this does not ``solve'' the cosmological
constant problem because we still have no means to
estimate $\LambdaUV$. Depending on the ultraviolet model,
it may be large.
But since an estimate of $\LambdaUV$
was never the aim of the sequester,
this criticism is unfair.
Instead, what has been achieved is that
\emph{if $\LambdaUV$ can somehow be made small,
its impact on the global spacetime geometry
is not destabilized by loops at much lower scales}.

Loosely speaking, this analysis shows that the
sequester is not a `field theory' mechanism,
in the sense that the properties
of loops are unmodified in the ultraviolet.
Rather, we have added a new form of matter $\Lambda$
that is constrained by its field equation
to cancel the portion of the vacuum energy sourced by
matter loops.
Ordinarily this would be of no benefit, because the energy density
associated with $\Lambda$ would itself gravitate.
As explained above,
the special feature of the action for $\Lambda$
is that its $\sigma$ part does not source any gravitational
field.
Heuristically, this allows us to `degravitate' or
`sequester' the vacuum energy by
storing it in $\sigma$.
When stored in this way the matter loops
are gravitationally inert.

After vacuum loops have been sequestered,
the effective source term
for the gravitational field
is the subtracted energy--momentum tensor
$\tau_{\mu\nu}$ computed in the metric $g_{\mu\nu}$,
together with a correction from its spacetime volume
average $\STaverage{{\tau^\rho}_\rho}$.
The size of this correction was estimated in
Refs.~\cite{Kaloper:2013zca,Kaloper:2014dqa},
who considered a model in which the unsequestered contribution
$\LambdaUV$ was set to zero.

Nevertheless, there is something surprising about this
outcome.
We are still working in the context of local field theory,
with its characteristic poor control of ultraviolet
effects.
Where has the original ultraviolet sensitivity
of the cosmological term gone?
The sequester \emph{does} contain a new physical ingredient,
in the form of the $\sigma$-term that is shielded
from gravity.
However, we have not introduced a new physical principle
that forces $\Lambda$ to capture
the entirety of $\LambdaSM$ in this non-gravitating sector.
Therefore one might worry that quantum corrections
`detune' the dynamical equation for $\Lambda$,
preventing complete sequestration
of $\LambdaSM$
and reintroducing the low-energy cosmological term.

\para{Radiative corrections}
Indeed,
when discussing any proposed solution to the
cosmological constant problem it is
\emph{never} sufficient
to work at tree level.
Like any naturalness problem,
the cosmological constant problem is intrinsically
quantum mechanical because it is only in a quantum theory
that loop corrections
generate direct correlations between widely separated
scales.
The conclusion is that radiative corrections must be included
before we can judge the merits of any
particular proposal.

A subset of relevant
corrections were considered
in Refs.~\cite{Kaloper:2013zca,Kaloper:2014dqa,Kaloper:2015jra,Kaloper:2016jsd}.
First, these authors considered a symmetry
$\lambda \rightarrow \Omega \lambda$,
$g_{\mu\nu} \rightarrow \Omega^{-2} g_{\mu\nu}$,
$\Lambda \rightarrow \Omega^4 \Lambda$
under which Eq.~\eqref{eq:model-one}
is invariant.
They argued this was sufficient to guarantee
that, to all orders in matter loops,
$\LambdaSM$ would couple to Eq.~\eqref{eq:model1-lambda-eqn}
like the tree-level vacuum energy.
In our presentation this symmetry is implied by
coupling $\Lmatter$ to the
Jordan-frame metric $\tilde{g}_{\mu\nu}$
but $\Lambda$ to the Einstein-frame
metric $g_{\mu\nu}$.
The loop-level behaviour of $\LambdaSM$
then follows from diffeomorphism invariance
with respect to $\tilde{g}_{\mu\nu}$.
We will give a pedestrian proof of these
properties in {\S}\ref{sec:model1-uv-sensitive}
below, based on analysis of
Feynman diagrams.
As we show there,
like all global symmetries, this one is broken
by coupling to gravity.

Second, Refs.~\cite{Kaloper:2013zca,Kaloper:2014dqa,Kaloper:2015jra,Kaloper:2016jsd}
studied the symmetry
$\Lambda \rightarrow \Lambda + \lambda^4 \nu^4$,
$\Lmatter \rightarrow \Lmatter - \nu^4$
which they suggested would guarantee
that $\Lambda$ absorbed
$\LambdaSM$ to all orders in matter loops.
[That is, that $\Lambda$ and $T_{\mu\nu}$ would appear
additively in the Einstein equation as in
Eq.~\eqref{eq:sequester-Einstein}.]
This last symmetry is not in fact a transformation
of the fields that participate in the
action, and is not respected by quantum
corrections.

\subsection{Ultraviolet sensitivity in Einstein frame}
\label{sec:model1-uv-sensitive}
This list does not exhaust the loop corrections to
Eq.~\eqref{eq:model-one}.
In particular,
the analysis of Refs.~\cite{Kaloper:2013zca,Kaloper:2014dqa,Kaloper:2015jra,Kaloper:2016jsd}
leaves open the issue of
(i)
corrections to the intensive global function
$\sigma$
that ``stores'' the unwanted large loop terms;
and
(ii)
corrections to the extensive
interaction $\int \d^4 x \, \sqrt{-g} \, \Lambda$.
To study corrections to $\sigma$
would require a microscopic theory that
explains how the flux $F_4$ is supported.
The sequester proposal does not aim to provide
such a description.
(For recent attempts to describe an ultraviolet
completion of this kind,
see Refs.~\cite{Padilla:2018hvp, Bordin:2019fek, El-Menoufi:2019qva, Sobral-Blanco:2020rdu, Alexander:2020tsf}.)
We comment on this in {\S}\ref{sec:conclusions}.
On the other hand,
the interaction term couples to spin-2
excitations of the metric $g_{\mu \nu}$,
and will therefore be ``dressed''
by loops containing off-shell quanta associated with
these excitations (cf. Ref.~\cite{Oda:2017bwx}).
This is a model independent effect, in the sense that
it does not depend on the microscopic origin of $F_4$.
In this section we aim to enumerate
these corrections and quantify
their impact.

\para{Loops respect diffeomorphism invariance}
First, we pause to prove the property stated above,
that pure matter loops
generate a cosmological term scaling as $\lambda^4$
to all orders in the loop expansion.
This follows from diffeomorphism invariance
with respect to $\tilde{g}_{\mu\nu}$, but can also be
proved by direct analysis of Feynman diagrams.
The results will assist us in an analysis of
corrections to the $\Lambda$ coupling, to be given below.

Consider any operator in $\Lmatter$
formed from a monomial of $n_b$ bosonic
fields and $n_f$ fermionic fields.
After replacing measured mass scales $\tilde{M}$
by their conformally rescaled equivalents
$M = \lambda \tilde{M}$,
and performing the same replacement
$k = \lambda \tilde{k}$
for momenta,
it can be checked that such an operator
scales like $\lambda^{n_b + 3n_f/2}$.
Meanwhile, a boson propagator scales
like $\lambda^{-2}$
whereas a fermion propagator scales
like $\lambda^{-3}$.
Therefore a diagram containing 
$I_b$ internal boson lines,
$E_b$ external boson lines,
$I_f$ internal fermion lines,
and $E_f$ external fermion lines
will scale like
$\lambda^D$, where
\begin{equation}
    D
    =
    -2 I_b -2 E_b - 3 I_f - 3 E_f
    +
    \sum_i N_i \Big( n_{b,i} + \frac{3}{2} n_{f,i} \Big)
    .
    \label{eq:lambda-scaling}
\end{equation}
$N_i$ is the number of vertices of type $i$,
each of which contains
$n_{b,i}$ bosonic fields and $n_{f,i}$ fermionic fields.

Each diagram must satisfy the
topological identity
$2I + E = \sum_k N_k n_k$,
where now $I$ denotes the total number of internal lines
(whether bosons or fermions),
$E$ denotes the total number of external lines,
$N_k$ denotes the number of vertices of type $k$,
and each type-$k$ vertex connects $n_k$ lines.
To translate to an operator in the effective action
we should amputate external lines.
Applying the identity separately to the bosonic and fermionic
components of the amputated diagram, it follows that
the effective operator will scale like
$\lambda^{\Damp}$, where
\begin{equation}
    \Damp = E_b + \frac{3}{2} E_f .
    \label{eq:feynman-lambda-scaling}
\end{equation}
(This analysis applies even if the bosonic and fermionic
components are disconnected,
provided the assignment of internal and external lines
is the one appropriate for the entire diagram.)
No matter how complex the diagram, Eq.~\eqref{eq:feynman-lambda-scaling}
involves only
the total number of amputated bosonic and fermionic lines.
Such a diagram will renormalize operators
that are polynomial in $E_b$ bosonic fields and
$E_f$ fermionic fields.
The $\lambda$-dependence of this renormalization
will be $\lambda^{E_b + 3 E_f/2}$, the same
as we deduced above for unrenormalized operators in $\Lmatter$.
The conclusion, as has already been stated,
is that pure matter loops preserve
the $\lambda$-dependence of the coupling in
Eq.~\eqref{eq:model-one}.%
    \footnote{Recall that this scaling applies
    \emph{after} conformal redefinition of the masses.
    From inspection of Eq.~\eqref{eq:model-one},
    one might expect the cosmological term generated
    by matter to scale as $\lambda^4 \tildeMSM^4$,
    where $\tildeMSM$ is some characteristic Standard
    Model scale.
    This does not conflict with~\eqref{eq:feynman-lambda-scaling}
    for $E_b = E_f = 0$
    because after rescaling
    $\MSM = \lambda \tildeMSM$
    the cosmological term scales as $\lambda^0$ as
    claimed.}

\para{Stability of global constraint}
Next, we argue that detuning the
dynamical equation for $\Lambda$
can prevent complete sequestration.
Specifically, to obtain
complete cancellation in the Einstein
equation, the factor of 4
that appears on the far left of Eq.~\eqref{eq:model1-lambda-eqn}
is required to match
a factor of $4$ from the trace of the metric in ${T^\mu}_\mu$.
Even a small mismatch of these factors will leave a residual
low-energy cosmological term in Eq.~\eqref{eq:sequester-Einstein}.

While the $4$ from the trace
${\delta^\mu}_\mu$ cannot be modified
by ultraviolet effects,
the other factor of $4$ is a consequence of the power $\lambda^{-4}$
appearing in the combination
$\Lambda/(\lambda \mu)^4$
that enters the global function $\sigma$.
We will argue below that this factor can be renormalized by
ultraviolet effects.
It follows that Eq.~\eqref{eq:model-one}
may receive significant corrections from
high energies,
and therefore fails the test for naturalness
in the sense we have defined.

How sensitive is the successful operation of the sequester
to the precise factor $4$ in Eq.~\eqref{eq:model1-lambda-eqn}?
If it is replaced by $4(1 + \alpha)$, the
analogue of the sequestered Einstein equation~\eqref{eq:model1-seq-Einstein-eqn}
becomes
\begin{equation}
\begin{split}
	\Mp^2 G_{\mu\nu}
	= \mbox{}
	&
	\frac{\alpha}{1+ \alpha} \lambda^4 \tildeLambdaSM g_{\mu\nu}
	+
	\tau_{\mu\nu}
	\\
	&
	{-
	\frac{1}{4(1+\alpha)}} \STaverage{{\tau^\mu}_\mu} g_{\mu\nu}
	+
	\LambdaUV g_{\mu\nu} .
\end{split}
\label{eq:alpha-Einstein-eq}
\end{equation}
As expected, there is now incomplete cancellation of
$\LambdaSM$.
To estimate the magnitude of the residual
cosmological term
requires
a numerical estimate for $\lambda$.
In a finite universe, Eq.~\eqref{eq:model1-Lambda-eqn}
yields
\begin{equation}
	\label{eq:model1-lambda}
	\lambda \sim \sigma' \frac{\Hage}{\mu} .
\end{equation}
The mass scale $\Hage$ was introduced
in Ref.~\cite{Kaloper:2014dqa}
and specifies the lifetime
of the universe.
This roughly determines the spacetime volume,
\begin{equation}
	\frac{1}{\Hage^4} \sim \int \d^4 x \; \sqrt{-g} .
	\label{eq:Hage-def}
\end{equation}
Clearly $\Hage < H_0 \sim 10^{-33} \, \eV$.

Ref.~\cite{Kaloper:2014dqa} suggested that
$\sigma$ should be engineered to obtain
$\lambda = \Or(1)$.
In this case, the effective gravitating
cosmological constant is
roughly
$\alpha \LambdaSM \sim \alpha \tildeLambdaSM \sim \alpha \tildemtop^4$,
assuming $|\alpha| \ll 1$.
With these estimates, $|\alpha|$
must inherit the tuning to 56 decimal
places that was previously required for
the combination
$\LambdaUV + \LambdaSM$.
If $\lambda$ is made smaller
then $\alpha$ can be relaxed
accordingly, but this scenario encounters
other difficulties~\cite{Kaloper:2014dqa}.

\para{Extensive corrections to the $\Lambda$ coupling}
Let us now estimate the model-independent
corrections to the extensive coupling
$-\Lambda V$, where
$V = \int \d^4 x \, \sqrt{-g}$.

\begin{figure}
    \begin{center}
        \includegraphics{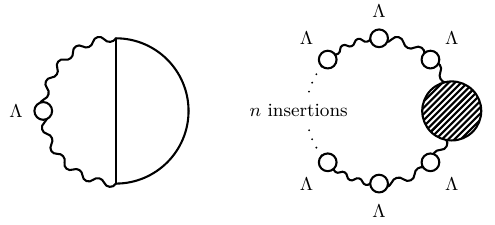}
    \end{center}
    \caption{\label{fig:Lambda-coupling-renormalize}Loops renormalizing the coupling of
    $\Lambda$ to the spacetime volume.
    \semibold{Left}: loop with single insertion of $\Lambda$ vertex,
    represented by the open circle. Wiggly lines represent spin-2
    excitations of the metric $g_{\mu\nu}$; solid lines represent Standard Model fields.
    This diagram renormalizes the coefficient of $\Lambda$ in $\sigma$.
    \semibold{Right}: loop containing $n$ insertions of $\Lambda$.
    This diagram renormalizes the coefficient of $\Lambda^{n}$ in $\sigma$.
    The shaded circle represents any Standard Model sub-diagram.
    The left-hand diagram is a particularly simple example of the class
    represented by the right-hand diagram.}
\end{figure}

First, consider the two-loop correction that appears in the left-hand
diagram of Fig.~\ref{fig:Lambda-coupling-renormalize}.
Regarded as a contribution to the quantum  effective
action, this contains a single insertion of a $\Lambda$ vertex
which is ``bridged'' to a pure Standard Model loop 
by a pair of spin-2 excitations.
This diagram is part of a larger class of diagrams,
represented by the right-hand part of
Fig.~\ref{fig:Lambda-coupling-renormalize},
in which an arbitrary number of $\Lambda$
insertions are bridged to a Standard Model
sub-diagram (of arbitrary complexity)
by graviton lines.

(These diagrams are not the only sources of renormalization
for the $\Lambda$ coupling.
We could equally well 
consider diagrams in which the $\Lambda$ insertions are embedded
within the Standard Model sub-diagram.
For our purpose, it suffices to consider
only a sub-class of possible renormalizations.)

According to the analysis given above,
the $\lambda$ dependence of this Standard Model
sub-diagram
can be computed from
Eq.~\eqref{eq:lambda-scaling}.
This time we are not amputating external lines, so the scaling is
$\lambda^{\Dsub}$ where $\Dsub = -E_b - 3E_f/2$.
Because the sub-diagram connects to the
ring of $\Lambda$ insertions via graviton lines
(which do not scale with $\lambda$) we have $E_b = E_f = 0$.
Meanwhile, counting the number of $\Lambda$ insertions
and a factor $\Mp^{-2}$ for each graviton propagator,
we conclude that such a diagram produces an operator
$\mathcal{O}$ in the
quantum effective action of the form
\begin{equation}
\begin{split}   
    \mathcal{O}_n
    & = 
    \frac{c_n}{(2\pi)^{4(L+1)}} \MSM^4 \frac{\Lambda}{\Mp^4} \left( \frac{\Lambda}{\Mp^2 \MSM^2} \right)^{n-1}
    \\
    & =
    \frac{c_n}{(2\pi)^{4(L+1)}} \Lambda^{n} \Mp^{-2(n+1)} \MSM^{6-2n} ,
\end{split}
\label{eq:lambda-scaling-estimate}
\end{equation}
where $L$ counts the number of loops in the Standard Model sub-diagram
and
$c_n$ is a Wilson coefficient that can be taken to be of order unity.
The scale $\MSM$ represents a typical Standard Model mass.
After replacing $\MSM$ by its experimentally-measurable counterpart
$\tildeMSM \sim \TeV$,
it follows that $\mathcal{O}$
scales like $\lambda^{6-2n}$.

To validate Eq.~\eqref{eq:lambda-scaling-estimate}
we have evaluated the explicit two-loop 
diagram given in the left-hand part of 
Fig.~\ref{fig:Lambda-coupling-renormalize},
for which $n=1$.
Using dimensional regularization as the ultraviolet
regulator,
this yields the expected scaling
\begin{equation}
    \mathcal{O}_{1}
    = \frac{c_1}{(2\pi)^8}
    \frac{\lambda^4 \tildeMSM^4}{\Mp^4}
    \Lambda .
    \label{eq:leading-correction}
\end{equation}
The factor $(2\pi)^{-8}$ is included from the
measure on the loop integrals.
Eq.~\eqref{eq:leading-correction}
will be the leading correction
provided $|\Lambda| \lesssim (\Mp \MSM)^{2}$.
This will generally be the case where $\Lambda$
is dynamically constrained to sequester
a loop contribution of order $\MSM^4$.
The field equations corrected by $\mathcal{O}_1$
are
\begin{subequations}
\begin{align}
    \label{eq:O1-Lambda-eqn}
    (1 - \lambda^4 \epsilon)
    \frac{\sigma'}{(\lambda\mu)^4}
    & =
    \int \d^4 x \; \sqrt{-g} ,
    \\
    \label{eq:O1-lambda-eqn}
    4 (1 + \lambda^4 \epsilon)
    \frac{\Lambda}{(\lambda \mu)^4} \sigma'
    & =
    \int \d^4 x \; \sqrt{-g} \; {T^\mu}_\mu ,
    \\
    \label{eq:O1-raw-Einstein-eqn}
    \Mp^2 G_{\mu\nu} & = T_{\mu\nu} -(1+\lambda^4 \epsilon) \Lambda g_{\mu\nu} + \LambdaUV g_{\mu\nu} ,
\end{align}
\end{subequations}
where $\epsilon \equiv (2\pi)^{-8} c_1 (\tildeMSM/\Mp)^4 \ll 1$
and we have dropped terms of $\Or(\epsilon^2)$.
After eliminating $\Lambda$, the Einstein equation can
be written, still up to $\Or(\epsilon)$ [cf.~\eqref{eq:alpha-Einstein-eq}],
\begin{equation}
    \Mp^2 G_{\mu\nu} = \Big( (\lambda^4 \epsilon) \LambdaSM + \LambdaUV \Big) g_{ab}
    + \tau_{\mu\nu} - \frac{1}{4} \STaverage{{\tau^\rho}_\rho} .
    \label{eq:O1-Einstein-sequestered}
\end{equation}
We have omitted $\Or(\epsilon)$ corrections
if they merely perturb existing
terms of order unity.
The conclusion is that
$\mathcal{O}_1$
corrects
Eqs.~\eqref{eq:O1-Lambda-eqn}--\eqref{eq:O1-lambda-eqn}
differently,
and therefore
renormalizes the relative factor $4$ between their left-hand
sides.

As in the analysis leading to Eq.~\eqref{eq:alpha-Einstein-eq},
this $\mathcal{O}_1$-corrected factor
no longer cancels the exact $4$
coming from ${\delta^\mu}_\mu$,
leaving a residual loop term in Eq.~\eqref{eq:O1-Einstein-sequestered}.
This outcome is practically inevitable.
In this formulation of the sequester, one is attempting
to balance a protected topological quantity ${\delta^\mu}_\mu$
against the properties of a class of unprotected
Lagrangian operators.
One is immune from ultraviolet effects but the other is not,
making the balance extremely delicate.

\para{Size of residual loop-level term}
How significant is this effect?
Taking $\lambda$ of order unity
and $\tildeMSM$ of order $1 \, \TeV$
makes $\epsilon$ of order $10^{-68}$,
or possibly as large as
$10^{-62}$ if we omit the loop-counting
factor $(2\pi)^{-8}$
on the assumption it is partially cancelled by
combinatorial factors.
Meanwhile if $\tildeLambdaSM$ is also of
order $1 \, \TeV$
then the residual loop-sourced cosmological
term in~\eqref{eq:O1-Einstein-sequestered}
is of order $(10^{-4} \, \eV)^4$ to
$(10^{-5} \, \eV)^4$,
or
$(10^{-4} \, \eV)^4$ to
$(10^{-3} \, \eV)^4$
if the loop-counting factor is omitted.
This is on the boundary of being acceptable given current
observational constraints.

The outcome is that whether the Einstein-frame
model can survive
ultraviolet corrections to the extensive coupling
is model-dependent.
Assuming the sequestered sector to be the Standard Model
gives a barely acceptable phenomenology,
with success or failure largely dependent on whether
$\lambda$ is larger
or smaller than unity.

Alternatively, if the sequestered sector contains
particles that are heavier than the Standard Model---for example,
perhaps from a higher-lying supersymmetric sector---%
then the model is unlikely to survive unless
$\lambda$ is significantly smaller than unity.
If the heaviest sequestered mass scale is
even $10 \, \TeV$
then the residual cosmological constant
is already in excess of the observed value.


\section{Jordan frame model}
\label{sec:sequester-model2}

\subsection{Graviton loop corrections}
\label{sec:graviton-loops}
The radiative corrections described
in~{\S}\ref{sec:model1-uv-sensitive}
above involved
loop diagrams containing virtual Einstein-frame
gravitons
that dress the $\Lambda$ coupling to the
spacetime volume.
There is a further class of diagrams of this type
that are significant in the Einstein frame. These are loop diagrams
containing virtual gravitons that
contribute to the low-energy cosmological constant.
As explained in Ref.~\cite{Kaloper:2016jsd},
and as we will review below,
these contributions escape the sequester.

Consider the left-hand diagram of Fig.~\ref{fig:cc-loops},
which is a loop diagram containing only matter fields.
As explained 
in the discussion leading to Eq.~\eqref{eq:feynman-lambda-scaling},
this diagram has $E_b = E_f = 0$ external lines
and therefore scales like $\lambda^0$
multiplied by $\MSM^4$.
Expressed in terms of the mass measured in a homogeneous gravitational
field this is $\propto \lambda^4 \tildeMSM$.

\para{Diagrams containing virtual gravitons}
Now consider the right-hand diagram of
Fig.~\ref{fig:cc-loops}.
In addition to matter fields (represented by the solid lines),
this contains an internal graviton
(represented by the wiggly line).
Each graviton propagator is proportional
to the fixed Planck scale $\Mp^{-2}$ with no
conformal rescaling.
It follows on dimensional grounds that
a renormalization of the cosmological
term with zero external lines,
any number of internal matter lines,
and $n_g$ internal graviton lines,
will scale as $\MSM^{4 + 2 n_g} / \Mp^{2n_g} \propto
\lambda^{4 + 2n_g}$.%
	\footnote{Such scalings are possibly modified by powers of logarithms,
	but we drop these unless they are dominant.}
If $n_g \neq 0$ these diagrams do not preserve
the $\lambda$ dependence of Eq.~\eqref{eq:model-one}~\cite{Kaloper:2016jsd}.

\subsection{Jordan-frame formulation}
\label{sec:model-two}
To solve this,
Ref.~\cite{Kaloper:2016jsd}
proposed an alternative description of the sequester
that we review below.
It is based on a reformulation of
Eq.~\eqref{eq:model-one} in the Jordan frame~\cite{Kaloper:2015jra}.
We wish to analyse the ultraviolet properties
of this formulation separately,
so we discuss it here before going on to consider
the problem of capturing diagrams containing virtual gravitons.

In Eq.~\eqref{eq:model-one}
the gravitational
action is built from $g_{\mu\nu}$,
but matter couples to $\tilde{g}_{\mu\nu}$.
The conformal factor between $g_{\mu\nu}$ and $\tilde{g}_{\mu\nu}$
adjusts the importance of
the matter action $\Lmatter$
relative to
the fixed Einstein term $R(g)$.
Alternatively, one can build the
action solely from the Jordan-frame metric,
leaving the relative importance of the Einstein term
as a free parameter,
\begin{equation}
\begin{split}
    S = \mbox{} & \int \d^4 x \, \; \sqrt{-g}
    \Bigg(
        \frac{\kappa^2}{2} R
        - \Lambda
        + \LambdaUV
        - \Lmatter    
    \Bigg)
    \\
    & \mbox{} 
    + \sigma\Big( \frac{\Lambda}{\mu^4} \Big)
    + \hat{\sigma}\Big( \frac{\kappa^2}{\Mp^2} \Big) .
\end{split}
\label{eq:model-two}
\end{equation}
This is the Jordan-frame formulation of the sequester.
The gravitational coupling is set by $\kappa$,
which is related to $\Mp$ by the global term
$\hat{\sigma}$.
As with $\sigma$, this should be a smooth function of
its argument
and
is assumed to be produced by integration
of a second flux, $\int \hat{F}_4$.
The Einstein frame metric $g_{\mu\nu}$ does not appear.

\para{Sequestration of low-energy loops}
The field equations that follow from~\eqref{eq:model-two}
are
\begin{subequations}
\begin{align}
	\label{eq:model2-einstein-eqn}
    \kappa^2 G_{\mu\nu} & =
    T_{\mu\nu} - \Big( \Lambda - \LambdaUV \Big) g_{\mu\nu} , \\
    \label{eq:model2-Lambda-eqn}
    \frac{1}{\mu^4} \sigma'\Big( \frac{\Lambda}{\mu^4} \Big) & =
    \int \d^4 x \, \sqrt{-g} , \\
    \label{eq:model2-kappa-eqn}
    \frac{1}{\Mp^2} \hat{\sigma}'\Big( \frac{\kappa^2}{\Mp^2} \Big) & =
    \mbox{} - \int \d^4 x \, \sqrt{-g} \; \frac{R}{2} .
\end{align}
\end{subequations}
Using the definition of spacetime average
$\STaverage{\cdots}$ given in {\S}\ref{sec:low-energy-model-one},
Eqs.~\eqref{eq:model2-Lambda-eqn}--\eqref{eq:model2-kappa-eqn}
require
\begin{equation}
	\label{eq:model2-Ravg-one}
	\STaverage{R} = -2 \frac{\mu^4}{\Mp^2} \frac{\hat{\sigma}'}{\sigma'} .
\end{equation}
Meanwhile, tracing the Einstein equation~\eqref{eq:model2-einstein-eqn}
and taking the spacetime average, we find
\begin{equation}
	\label{eq:model2-Ravg-two}
	\STaverage{R} = - \frac{1}{\kappa^2}
	\left[
		\STaverage{{T^\mu}_\mu}
		- 4 ( \Lambda - \LambdaUV )
	\right] .
\end{equation}
Eqs.~\eqref{eq:model2-Ravg-one} and~\eqref{eq:model2-Ravg-two}
must hold simultaneously, and therefore
\begin{equation}
	\Lambda - \LambdaUV = \frac{1}{4} \STaverage{{T^\mu}_\mu}
	- \frac{\mu^4}{2} \frac{\kappa^2}{\Mp^2} \frac{\hat{\sigma}'}{\sigma'} .	
	\label{eq:model2-Lambda-soln}
\end{equation}
Finally, we replace $\Lambda$ in the Einstein equation to obtain
\begin{equation}
	\kappa^2 G_{\mu\nu} = T_{\mu\nu} - \frac{1}{4} \STaverage{{T^\mu}_\mu}
	+ \frac{\mu^4}{2} \frac{\kappa^2}{\Mp^2} \frac{\hat{\sigma}'}{\sigma'} 
	g_{\mu\nu}.
	\label{eq:model2-seq-Einstein-eqn}
\end{equation}

\para{Relation between Einstein- and Jordan-frame}
The Einstein- and Jordan-frame formulations are related by a change
of frame, and therefore must presumably
be regarded as equivalent.
This equivalence holds even
up to quantum corrections provided
one is sufficiently careful to include contributions
from the transformation Jacobian; see, e.g. Ref.~\cite{Brax:2010uq}.
The key issue to be addressed is how ultraviolet modes
enter each formulation,
to be discussed in {\S}\ref{sec:model1-uv-sensitive}.

Before doing so, we enumerate the principal
differences between the sequester phenomenology
in Einstein frame and Jordan frame.
First, in Jordan frame, not only the low-energy loop contribution $\LambdaSM$
is sequestered, but also the ultraviolet part $\LambdaUV$.
This happens because both sources for the cosmological
term now couple to the Jordan-frame metric.
The distinction between them is therefore arbitrary at the level
of the Einstein equation.
We will see below that this emerges from a more general conclusion,
that fluctuations coupling to the Jordan-frame metric
(including gravitons)
are sequestered, whereas fluctuations coupling to
the Einstein-frame metric are not.

Second, the critical factor of $1/4$ in the combination
$T_{\mu\nu} - \STaverage{{T^\mu}_\mu}/4$
is \emph{not ultraviolet sensitive}.
In particular, it is no longer produced by balancing a topological
invariant against the properties of a particular
group of Lagrangian operators.
Instead, the factor of $1/4$ in Eqs.~\eqref{eq:model2-Lambda-eqn}
and~\eqref{eq:model2-seq-Einstein-eqn}
is also produced by a trace.
Therefore it is not corrected by extensive renormalizations
of the coupling of $\Lambda$ to spacetime.
We will consider below what is the effect of these
renormalizations in the Jordan frame.

Third, the Jordan frame formulation generates
a residual cosmological-like term.
This is the last term
in~\eqref{eq:model2-seq-Einstein-eqn}.
Assuming $\kappa^2 \sim \Mp^2$
and $\sigma' \sim \hat{\sigma}' \sim \Or(1)$,
it yields a residual cosmological constant of order
$\mu^4$.
Ref.~\cite{Kaloper:2016jsd} argued that
this contribution is at least radiatively stable
because it arises from the intensive term $\sigma$,
which
does not couple either to $g_{ab}$ or
the matter fields in $\Lmatter$.
It is therefore uncorrected by matter and graviton
loops.
On the other hand, depending on its origin,
$\sigma$ might be susceptible to other loop corrections
associated with unknown mass scales.
If so, $\mu$ must apparently be associated with the
\emph{lowest} of these scales,
because it is the most relevant
terms involving $\Lambda$ that
dominate Eq.~\eqref{eq:model2-Ravg-two}.
(However, it should be remembered that $\mu$ does not
have a precise meaning until we specify the typical
size of Taylor coefficients in $\sigma$.)

This does not preclude the possibility
that $\mu$ could typically be large.
As with the cosmological constant itself
this need not be fatal for the model,
because we can always suppose that
the renormalized value of $\mu$ is
lower than its natural scale.
If we choose to do so, however, then
presumably we encounter a new naturalness
problem in the
$\Lambda$ sector.
In particular, $\Lambda$ must become relevant at a
very low energy scale $\sim (10^{-3} \, \eV)^4$
to avoid an unwanted large contribution.

At the level of the effective theory~\eqref{eq:model-two}
there is nothing further that can be said
to set our expectations about the typical size of $\mu$.
To do so would require a detailed microscopic theory of the
fluxes and how they are sourced.
Such a theory could be used to compute corrections to
the functions $\sigma$ and $\hat{\sigma}$.
In this connection, see Refs.~\cite{Padilla:2018hvp,El-Menoufi:2019qva}.

\subsection{Ultraviolet dependence in Jordan frame}
In Eq.~\eqref{eq:model-two}
there will be extensive renormalizations of the
coupling of
$\kappa$ and $\Lambda$ to spacetime.
Note that $\Lambda$ couples to the spacetime volume,
whereas $\kappa^2$ couples to the integrated curvature
$\int \d^4 x \, \sqrt{-g} \, R$.
Renormalizations of the $\Lambda$ coupling were
considered above and are unchanged in this theory.
Renormalization of the $\kappa$ coupling will arise
from diagrams analogous to those of Fig.~\ref{fig:Lambda-coupling-renormalize},
but with insertions of $\kappa^2 R$ rather than $\Lambda$.
(As before, the class of diagrams shown on the right-hand side
of Fig.~\ref{fig:Lambda-coupling-renormalize}
does not exhaust the contributions at a given order in $\kappa^2 R$,
but they provide a representative class that is simple to study.)
The leading effects can be summarized by the replacements
\begin{subequations}
\begin{align}
    \frac{\kappa^2}{2} R
    & \rightarrow
    \frac{\kappa^2}{2} ( 1 + \alpha \epsilon ) R ,
    \\
    \label{eq:model2-Lambda-dressing}
    \Lambda
    & \rightarrow
    (1 + \beta \epsilon ) \Lambda
\end{align}    
\end{subequations}
where $\alpha$ and $\beta$ are $\Or(1)$ Wilson coefficients,
and $\epsilon$ is defined by
\begin{equation}
    \label{eq:model2-epsilon-def}
    \epsilon \equiv \frac{1}{(2\pi)^8} \frac{\MSM^4}{\kappa^4} \ll 1 .    
\end{equation}
To account for renormalizations of the low-energy
cosmological constant from diagrams including virtual gravitons,
as in the right-hand diagram of
Fig.~\ref{fig:cc-loops},
we include a representative term $\gamma \MSM^6 / \kappa^2$ in $\Lmatter$,
where $\gamma$ is another $\Or(1)$ coefficient.
A contribution of this form will be generated by diagrams
such as the right-hand side of Fig.~\ref{fig:cc-loops}
containing a single internal graviton line.
It would typically be accompanied by contributions of
higher order in $\kappa^{-2}$
from diagrams containing two or more internal graviton lines,
but if the scale $\MSM$ of the sequestered sector
is far below the Planck scale then the one-graviton
diagram will be dominant.
Notice that this term will contribute to the $\kappa$
field equation. This is the origin of the mismatch
that allows such contributions to escape complete sequestration.

After a short calculation, it follows that
the effective Einstein equation in this model can be written,
up to $\Or(\epsilon)$,
\begin{equation}
\begin{split}
    \kappa^2 G_{\mu\nu}
    = \mbox{}
    &
    (1-\alpha \epsilon) T_{\mu\nu} -
    \Big( 1 - (\alpha + \beta) \epsilon \Big) \frac{\STaverage{{T^\rho}_\rho}}{4} g_{\mu\nu}
    \\
    &
    \mbox{} +
    \beta \epsilon \LambdaUV g_{\mu\nu}
    -
    \frac{\gamma}{2} \frac{\MSM^6}{\kappa^2}
    \\
    &
    \mbox{}
    + \frac{\mu^4}{2} \frac{\kappa^2}{\Mp^2} \frac{\hat{\sigma}'}{\sigma'}
    ( 1 + \alpha \epsilon ) g_{\mu\nu} .
\end{split}
\label{eq:model2-uv-Einstein-eqn}
\end{equation}
We can identify a number of effects.
First, dressing of the $\kappa^2$ coupling
(proportional to $\alpha)$
can be absorbed into a redefinition of the Planck scale.
It does not cause de-tuning of the sequester.
By comparison, we cannot simply absorb~\eqref{eq:model2-Lambda-dressing}
into a redefinition of $\Lambda$ because of its $\kappa$
dependence.

Second, dressing of the $\Lambda$ coupling
(proportional to $\beta$)
is again responsible for breaking complete
cancellation of the low-energy cosmological contribution
between $T_{\mu\nu}$ and $\STaverage{{T^\mu}_\mu}/4$.
The residual cosmological constant
will be of order $\epsilon \LambdaSM \sim \epsilon \MSM^4$
and therefore of a similar size to the estimates
for the Einstein frame
given at the end of {\S}\ref{sec:model1-uv-sensitive}.
For numerical values we refer to the discussion given there.

Third, the $\Lambda$ dressing \emph{also}
causes inexact cancellation
of the ultraviolet part $\LambdaUV$.
The leftover piece has exactly the same structure
as the left-over low energy loop contribution in $T_{\mu\nu}$,
again because there is no distinction between these terms
at the level of the Einstein equation.
In the remainder of this paper
we shall drop explicit dependence on $\LambdaUV$
and include its contribution in $\Lmatter$ if required.
Finally, we clearly see the contribution of the
right-hand diagram in Fig.~\ref{fig:cc-loops};
this produces the term proportional to
$\gamma \MSM^6 / \kappa^2$~\cite{Kaloper:2016jsd}.

\subsection{Sequestering the graviton loops}
\label{sec:omnia-sequestra}

Kaloper et al. observed that the
troublesome
$\gamma$ term
appears in Eq.~\eqref{eq:model2-uv-Einstein-eqn}
as a consequence of
its appearance in the $\kappa$ field equation~\cite{Kaloper:2016jsd}.
If it could be removed from this field
equation then terms of any order in $\kappa^{-2}$
contained in $T_{\mu\nu}$
would be sequestered
as part of the usual cancellation between
$T_{\mu\nu}$ and $\STaverage{{T^\mu}_\mu}/4$,
at least in the absence of renormalizations to
the $\Lambda$ coupling to the volume of spacetime.

In turn, graviton-loop contributions to the low-energy
cosmological term
contribute to the $\kappa$ field equation \emph{only}
because the graviton
propagator carries a normalization of $\kappa^{-2}$.
To decouple these contributions
Kaloper et al. proposed the following formulation
(which they described as `omnia sequestra')~\cite{Kaloper:2016jsd,Coltman:2019mql}
\begin{equation}
\begin{split}
    S = & \int \d^4 x \; \sqrt{-g} \Bigg( \frac{\Mp^2}{2} R
    + \theta \RGaussBonnet - \Lambda
    - \Lmatter \Bigg)
    \\ &
    \mbox{} + \sigma\left( \frac{\Lambda}{\mu^4} \right)
    + \hat{\sigma}(\theta) .
\end{split}
\label{eq:omnia-sequestra}
\end{equation}
Recall that we are now absorbing $\LambdaUV$, if present, into $\Lmatter$.
The normalization of the Einstein term reverts to the fixed
Planck scale $\Mp$.
Meanwhile we introduce the Gauss--Bonnet density $\RGaussBonnet$
coupled to a rigid scalar $\theta$
that replaces $\kappa$.
The Gauss--Bonnet density is defined by
\begin{equation}
    \RGaussBonnet \equiv R^2 - 4 R^{\mu\nu} R_{\mu\nu} + R^{\mu\nu\rho\sigma} R_{\mu\nu\rho\sigma} .    
\end{equation}
In four dimensions
its integral is proportional to a topological invariant,
the Euler characteristic $\chi(M)$ of the manifold $M$.
Because it is topological (it integrates to a boundary term),
it follows that $\RGaussBonnet$
does not modify the form of the graviton propagator
or its self-interaction vertices.
The conclusion is that each internal graviton line scales
like $\Mp^{-2}$ and carries no $\theta$ dependence.
Operators in the quantum effective 
action that are built from diagrams containing such
lines do not perturb the field equation for
$\theta$.

Further,
because of its topological character,
the coefficient of the Gauss--Bonnet density
is not renormalized.
At the level of Feynman diagrams this follows because
$\theta$ does not contribute to graviton vertices.
Therefore there is no analogue of the diagrams in
Fig.~\ref{fig:Lambda-coupling-renormalize} for
$\RGaussBonnet$.
For the same reason, quantum corrections do not introduce $\theta$-dependence
in $\Lmatter$ at any order in the loop expansion.

The coupling of $\Lambda$ to the spacetime volume
will still be dressed by graviton loops, yielding 
Eq.~\eqref{eq:model2-Lambda-dressing}
with the replacement $\kappa^2 \rightarrow \Mp^2$
in $\epsilon$.
However, unlike Eqs.~\eqref{eq:model2-Lambda-dressing}
and~\eqref{eq:model2-epsilon-def},
there is now no obstruction to absorbing
the loop correction into a redefinition of $\Lambda$.
Accordingly we do not expect de-tuning of the sequester
in this case.

To verify this expectation consider the field equations
following from~\eqref{eq:omnia-sequestra}
with the leading loop correction to the $\Lambda$ coupling
included,
\begin{subequations}
\begin{align}
    \label{eq:model3-raw-Einstein-eqn}
    \Mp^2 G_{\mu\nu} 
    & =
    T_{\mu\nu} - (1 + \alpha \epsilon) \Lambda g_{\mu\nu} ,
    \\
    \label{eq:model3-Lambda-eqn}
    \frac{\sigma'}{\mu^4}
    & =
    ( 1+ \alpha \epsilon) \int \d^4 x \; \sqrt{-g} 
    \\
    \label{eq:model3-theta-eqn}
    \hat{\sigma}'
    & =
    - \int \d^4 x \; \sqrt{-g} \; \RGaussBonnet
\end{align}
\end{subequations}
From Eqs.~\eqref{eq:model3-Lambda-eqn}--\eqref{eq:model3-theta-eqn}
we conclude
\begin{equation}
    \frac{\hat{\sigma}'}{\sigma'} \mu^4 = -(1-\alpha\epsilon) \STaverage{\RGaussBonnet} .
    \label{eq:GaussBonnet-mu-estimate}
\end{equation}
Because the Gauss--Bonnet density integrates to the
Euler characteristic, up to a numerical factor,
this is a relatively stringent condition on $\mu$.
Assuming the derivatives $\sigma'$ and $\hat{\sigma}'$ are order unity%
    \footnote{In our presentation, we are absorbing the
    integrated fluxes $\int F_4$, $\int \hat{F}_4$
    into the definition of $\sigma$, $\hat{\sigma}$.
    If these factors are large
    they may modify conclusions based on
    dimensional analysis of~\eqref{eq:GaussBonnet-mu-estimate}.}
it roughly requires $\mu \sim \Hage$, where $\Hage$ is the
quantity defined in~\eqref{eq:Hage-def}.
See also Ref.~\cite{Coltman:2019mql}.

Meanwhile, the trace of the Einstein equations
requires
\begin{equation}
    R = \frac{4}{\Mp^2} (1 + \alpha \epsilon) \Lambda - \frac{1}{\Mp^2} {T^\mu}_\mu .    
\end{equation}
As in the analyses given above,
taking the spacetime expectation of
this formula gives an expression for
$\Lambda$ in terms of $\STaverage{R}$ and $\STaverage{{T^\mu}_\mu}$.
This expression should be used to eliminate $\Lambda$ from the Einstein equation.
Finally, expressing $\STaverage{R}$ in terms of $\STaverage{\RGaussBonnet}$
yields
\begin{equation}
    \label{eq:omnia-sequestra-Einstein-eqn}
    \Mp^2 G_{\mu\nu} = T_{\mu\nu} - \frac{1}{4} \STaverage{{T^\rho}_\rho} g_{\mu\nu}
    - L g_{\mu\nu} ,
\end{equation}
where $\ell$ is defined by (cf. Eqs.~(11)--(12) of Ref.~\cite{Kaloper:2016jsd})
\begin{equation}
\label{eq:omnia-sequestra-residual}
\begin{split}
    L^2 = \frac{3}{8} \Mp^4 \Bigg(
        &
        \STaverage{\RGaussBonnet}
        - \STaverage{W^2}
        + \frac{2}{\Mp^2} \STaverage{(T_{\mu\nu} - T g_{\mu\nu}/4)^2}
    \\
        & \mbox{}
        - \frac{1}{6 \Mp^4} \Big[ \STaverage{T^2} - \STaverage{T}^2 \Big]
    \Bigg) ,
\end{split}
\end{equation}
where $T = {T^\rho}_\rho$
and $W_{\mu\nu\rho\sigma}$ is the Weyl tensor derived from $g_{\mu\nu}$.
This is exactly the result derived in Ref.~\cite{Kaloper:2016jsd}.
As expected, dressing of the $\Lambda$ coupling has no effect at the level
of the effective Einstein equation.
We conclude that extensive renormalizations of the
coupling between $\Lambda$ and the spacetime volume
do \emph{not} de-tune sequestration in the formulation~\eqref{eq:omnia-sequestra}.

\section{Conclusions}
\label{sec:conclusions}
In this paper we have studied a class of radiative corrections to the 
sequester model proposed by Kaloper, Padilla and collaborators.
Although the corrections we compute have previously been recognized, their
effect has not been studied explicitly.
The class of diagrams we study renormalize the couplings between
the ``rigid'' scalar fields that are characteristic of the sequester
scenario,
and infrared properties of the spacetime such as its volume and integrated
curvature.

In both the Einstein and Jordan frame formulations (given by Eqs.~\eqref{eq:model-one}
and~\eqref{eq:model-two} in our notation),
we find that these renormalizations disrupt complete sequestration of
low-energy loop contributions.
If the sequestered sector is the Standard Model, we find that these
corrections very nearly produce an unacceptable cosmological
term in excess of the observed value $\Lambda \sim (10^{-3} \, \eV)^4$.
Whether or not a particular realization of the scenario yields
an acceptable phenomenology then depends on how the
global function $\sigma$ is engineered
(and likewise for $\hat{\sigma}$ in the Jordan-frame formulation).

Alternatively, if the sequestered sector contains higher mass
particles such as supersymmetric partners with masses in excess
of $10 \, \TeV$, the residual cosmological term is likely to be
fatal. The situation could possibly be saved
if physical scales are significantly rescaled
in the effective Einstein frame metric.
This is easiest to see in the explicit Einstein-frame
description, where masses are rescaled by the
conformal factor $\lambda$.
We can possibly arrange for this rescaling $\lambda$ to be small,
but such scenarios encounter other difficulties~\cite{Kaloper:2014dqa}.

The simpler formulations of the sequester
(those that do not invoke the Gauss--Bonnet density)
are already known to ``fail''
in the sense that they do not capture contributions to the vacuum
energy from diagrams that contain virtual gravitons.
Although the renormalizations we have computed are related to
these known failure modes, they are not the same.
In most models the
loop terms we compute are likely to be somewhat smaller, since the
leading contribution involves two virtual gravitons
and therefore scale
as $(\MSM/\Mp)^4$.
This should be
compared to
a single-graviton loop scaling
as $(\MSM/\Mp)^2$ as in the left-hand diagram of Fig.~\ref{fig:cc-loops}.

We find that these renormalizations do not affect
the most developed formulation of the sequester,
given by Eq.~\eqref{eq:omnia-sequestra} in our notation.
In this formulation, dressing of the $\Lambda$ interaction can be
absorbed into a redefinition of $\Lambda$ itself and is therefore
harmless.

Whether or not one finds the sequester a plausible
solution to the naturalness problem of the cosmological
constant depends on whether we are prepared to accept
its key ingredient---the introduction of
non-gravitating sectors in the action
that are shielded from gravity:
they do not source gravitational fields,
and they do not interact with gravitons.
For related models utilising a similar
premise see Refs.~\cite{Oda:2017qce,Carroll:2017gqo,Sobral-Blanco:2020rdu,Bordin:2019fek}.
This is the cost of entry for all versions of the
sequester scenario.
Once accepted, it is only necessary to arrange for
the large low-energy loop contribution to be `stored'
in these non-gravitating sectors.

At the level of the effective actions used in this paper
there is little more that can be said.
In particular, we have not been able to apply
``naturalness'' arguments to the non-gravitating
functions $\sigma$ and $\hat{\sigma}$,
because to do so would require specification
of a microscopic theory that describes the
fluxes $F_4$, $\hat{F}_4$ that project out
local degrees of freedom from the rigid
fields $\Lambda$, $\kappa$ and $\theta$
(depending on the formulation in use).
These non-gravitating sectors
are the final repository for sequestered
vacuum energy.
If it is possible to build models in which
these sectors have their own microscopic description,
it would be very interesting
to apply naturalness criteria to the model as a whole.

\begin{acknowledgments}
MGL acknowledges support from the UK Science and Technology
Facilities Council via Research Training Grant
ST/M503836/1.    
DS acknowledges support from the Science and Technology Facilities Council
[grant number ST/L000652/1].
We would like to thank Chris Byrnes for helpful discussions throughout
this project.
We thank Chris Byrnes and Tony Padilla
for helpful comments on an early version of this
paper.
\end{acknowledgments}
\end{fmffile}

\bibliography{refs}
\end{document}